\documentstyle[twoside,12pt,epsf]{article}
\input epsf
\epsfverbosetrue

\def\singlespace {\smallskipamount=3pt plus1pt minus1pt
                  \medskipamount=6pt plus2pt minus2pt
                  \bigskipamount=12pt plus4pt minus4pt
                  \normalbaselineskip=12pt plus0pt minus0pt
                  \normallineskip=1pt
                  \normallineskiplimit=0pt
                  \jot=3pt
                  {\def\smallskip {\vskip\smallskipamount}}
                  {\def\medskip   {\vskip\medskipamount}}
                  {\def\bigskip   {\vskip\bigskipamount}}
                  {\setbox\strutbox=\hbox{\vrule
                    height8.5pt depth3.5pt width 0pt}}
                  \parskip 0pt
                  \normalbaselines}
\def\doublespace {\smallskipamount=6pt plus2pt minus2pt
                  \medskipamount=12pt plus4pt minus4pt
                  \bigskipamount=24pt plus8pt minus8pt
                  \normalbaselineskip=24pt plus0pt minus0pt
                  \normallineskip=2pt
                  \normallineskiplimit=0pt
                  \jot=6pt
                  {\def\smallskip {\vskip\smallskipamount}}
                  {\def\medskip   {\vskip\medskipamount}}
                  {\def\bigskip   {\vskip\bigskipamount}}
                  {\setbox\strutbox=\hbox{\vrule
                    height17.0pt depth7.0pt width 0pt}}
                  \parskip 12.0pt
                  \normalbaselines}
\def\halfspace {\smallskipamount=6pt plus2pt minus2pt
                  \medskipamount=12pt plus4pt minus4pt
                  \bigskipamount=24pt plus8pt minus8pt
                  \normalbaselineskip=16pt plus0pt minus0pt
                  \normallineskip=2pt
                  \normallineskiplimit=0pt
                  \jot=6pt
                  {\def\smallskip {\vskip\smallskipamount}}
                  {\def\medskip   {\vskip\medskipamount}}
                  {\def\bigskip   {\vskip\bigskipamount}}
                  {\setbox\strutbox=\hbox{\vrule
                    height17.0pt depth7.0pt width 0pt}}
                  \parskip 12.0pt
                  \normalbaselines}

\def\pprintspace {\smallskipamount=4pt plus1pt minus1pt
                  \medskipamount=9pt plus2pt minus2pt
                  \bigskipamount=16pt plus4pt minus4pt
                  \normalbaselineskip=14pt plus0pt minus0pt
                  \normallineskip=1pt
                  \normallineskiplimit=0pt
                  \jot=4pt
                  {\def\smallskip {\vskip\smallskipamount}}
                  {\def\medskip   {\vskip\medskipamount}}
                  {\def\bigskip   {\vskip\bigskipamount}}
                  {\setbox\strutbox=\hbox{\vrule
                   height9.5pt depth4.5pt width 0pt}}
                  \parskip 0pt
                  \normalbaselines}
\def\reidelspace {\smallskipamount=.1667 true in plus4pt minus2pt
                  \medskipamount=.3333 true in plus8pt minus2pt
                  \bigskipamount=13 true pt plus2pt minus2pt
                  \normalbaselineskip=13 true pt plus0pt minus0pt
                  \normallineskip=1 true pt
                  \normallineskiplimit=0 true pt
                  \jot=3pt
                  {\def\smallskip {\vskip\smallskipamount}}
                  {\def\medskip   {\vskip\medskipamount}}
                  {\def\bigskip   {\vskip\bigskipamount}}
                  {\setbox\strutbox=\hbox{\vrule
                    height8.5pt depth3.5pt width 0pt}}
                  \parskip 0pt
                  \normalbaselines}

\def\folio{\ifnum\pageno=1\nopagenumbers\else\number\pageno\fi}

\def\MeanQ  {10.7}		
\def\dQStat {3.6}		
\def\dQGal  {7.1}		

\def\refitem{\par\noindent\hangindent 20pt}
\def\wisk#1{\ifmmode{#1}\else{$#1$}\fi}
\def\lt     {\wisk{<}}

\def\ge     {\wisk{_>\atop^=}}

\def\um     {\wisk{{\rm \mu m}}}
\def\muK    {\wisk{{\rm \mu K}}}

\def\deg    {\wisk{^\circ}}

\def\thin{\thinspace}

\begin{document}
\pagestyle{plain}
\pprintspace

\large
\begin{center}
Microwave Emission at High Galactic Latitudes
\end{center}

\medskip
\normalsize
\pprintspace
\noindent
\begin{center}
A.~Kogut\footnotemark[1]$^{,2}$,
G. Hinshaw$^1$,
A.J. Banday$^{1,3}$,
C.L. Bennett$^4$,
K. G\'{o}rski$^{1,5}$,
G.F. Smoot$^6$,
and
E.L. Wright$^7$
\end{center}
\footnotetext[1]{
~Hughes STX Corporation, Laboratory for Astronomy and Solar Physics, 
Code 685, NASA/GSFC, Greenbelt MD 20771. \newline
\indent~$^2$ E-mail: kogut@stars.gsfc.nasa.gov. \newline
\indent~$^3$ Current address: Max Planck Institut f\"{u}r Astrophysik,
85740 Garching Bei M\"{u}nchen, Germany. \newline
\indent~$^4$ Laboratory for Astronomy and Solar Physics, 
Code 685, NASA/GSFC, Greenbelt MD 20771. \newline
\indent~$^5$ On leave from Warsaw University Observatory,
Aleje Ujazdowskie 4, 00-478 Warszawa, Poland. \newline
\indent~$^6$ LBL, SSL, \& CfPA, Bldg 50-25, University of California,
Berkeley, CA 94720. \newline
\indent~$^7$ UCLA Astronomy, PO Box 951562, Los Angeles, CA 90095-1562. \newline
}

\medskip
\normalsize
\pprintspace
\begin{center}
{\it COBE} Preprint 96-02 \\
Submitted to {\it The Astrophysical Journal Letters} \\
January 5, 1996 \\
\end{center}


\medskip
\begin{center}
\large
ABSTRACT
\end{center}

\normalsize
\noindent
We use the
{\it COBE\thin}\footnotemark[8]
\noindent \footnotetext[8]{
~The National Aeronautics and Space Administration/Goddard Space Flight Center 
(NASA/GSFC) is responsible for the design, development, and operation of the
Cosmic Background Explorer ({\it COBE}).  
Scientific guidance is provided by the {\it COBE} Science Working Group.  
GSFC is also responsible for the analysis software 
and for the production of the mission data sets.
}
Differential Microwave Radiometers (DMR) 4-year sky maps 
to model Galactic microwave emission at high latitudes ($|b| > 20\deg$).
Cross-correlation of the DMR maps with Galactic template maps
detects fluctuations in the high-latitude microwave sky brightness 
with the angular variation of the DIRBE far-infrared dust maps and
a frequency dependence consistent with a superposition of dust and free-free
emission.  We find no significant correlations between the DMR maps
and various synchrotron templates.
On the largest angular scales (e.g., quadrupole),
Galactic emission is comparable in amplitude to the anisotropy in the
cosmic microwave background (CMB).
The CMB quadrupole amplitude,
after correction for Galactic emission,
has amplitude $Q_{rms}$ = \MeanQ ~\muK\
with random uncertainty \dQStat ~\muK\
and systematic uncertainty \dQGal ~\muK\ from 
uncertainty in our knowledge of Galactic microwave emission.

{\it Subject headings:} 
cosmic microwave background --
cosmology: observations --
Galaxy : general --
ISM: general

\clearpage
\section{Introduction}

Diffuse microwave emission at high Galactic latitudes is dominated by the
cosmic microwave background
and optically thin emission from Galactic
synchrotron, dust, and free-free emission.
These components may be distinguished by their different
spatial morphology and frequency dependence.
A number of authors have attempted to separate Galactic and cosmic
emission on angular scales above a few degrees
(Fixsen, Cheng, \& Wilkinson 1983;
Lubin et al.\ 1985;
Wright et al.\ 1991;
Bennett et al.\ 1992;
Bensadoun et al.\ 1993;
Guti\'{e}rrez de la Cruz et al.\ 1995;
Kogut et al.\ 1996a).
Unfortunately, there is currently no emission component
for which both the spatial template and frequency dependence are well 
determined.
Synchrotron radiation dominates radio-frequency surveys,
but the spectral index $\beta_{\rm synch}$ 
steepens with frequency and
has poorly-determined spatial variation 
(Banday \& Wolfendale 1991, Bennett et al.\ 1992).
Dust emission dominates far-infrared surveys,
but its spectral behavior at longer wavelengths
depends on the shape, composition, and size distribution
of the dust grains, which are poorly known
(D\'{e}sert, Boulanger, \& Puget 1990).
Free-free emission from electron-ion interactions 
has well-determined spectral behavior
but lacks an obvious template map:
free-free emission never dominates the high-latitude radio sky,
while other tracers of the warm ionized interstellar medium (WIM)
such as H$\alpha$ emission, pulsar dispersion measure, 
or N{\tenrm II} emission
are either incomplete, undersampled, or noise-dominated
(Reynolds 1992,
Bennett et al.\ 1992, 1994).
At least one component of the WIM is spatially correlated with the 
far-infrared dust distribution on large angular scales (Kogut et al.\ 1996a); 
however, the fraction of the total WIM contained in the correlated component
has substantial uncertainties.

The ratio of cosmic to Galactic emission depends on the angular scale
and observing frequency.
CMB anisotropies have antenna temperature\footnotemark[9]
\footnotetext[9]{
~Antenna temperature $T_A$ is defined in terms of the power received per unit
bandwidth, $P = k T_A \Delta \nu$ where $k$ is Boltzmann's constant.
It is related to the intensity $I_\nu$ by
$I_\nu = 2 k T_A \frac{\nu^2}{c^2}$.
}
$$
\Delta T_A ~= ~\frac{ x^2 e^x }{(e^x - 1)^2} ~\Delta T
$$
where
$x = h\nu / kT$
and $T$ is thermodynamic temperature.
The angular power spectrum of CMB anisotropy 
is well described by a power-law spectrum
$P_{\rm CMB} \propto \frac{1}{\ell (\ell + 1)}$
(G\'{o}rski et al.\ 1996),
while Galactic emission follows a steeper law
$P_{\rm Gal} \propto \ell^{-3}$
(Gautier et al.\ 1992,
Kogut et al.\ 1996a),
where $\ell \propto \theta^{-1}$ is the spherical harmonic multipole order.
Galactic emission reaches a minimum near 60 GHz.
In this Letter we derive models of Galactic emission based on
Galaxy-dominated sky surveys
and the 4-year {\it COBE} DMR microwave maps.

\section{Techniques}
We use three main techniques to identify Galactic emission
in the high-latitude portion of the DMR maps,
here defined as the region $|b| > 20\deg$ 
with custom cutouts at Orion and Ophiuchus
(Bennett et al.\ 1996).
A ``subtraction'' method 
(Bennett et al.\ 1992)
scales Galactic template maps to the DMR frequencies
using spectral indices fixed by external observations.
A ``linear combination'' method corrects the DMR maps for
synchrotron and dust emission using the subtraction technique, then
fits the corrected maps pixel by pixel
for the CMB and free-free amplitude
(the components whose spatial distribution is least known).
A ``cross-correlation'' method cross-correlates the DMR maps with
fixed Galactic template maps without specifying any
{\it a priori} frequency dependence.
Each method has certain deficiencies.
The subtraction method suffers from uncertainty in the frequency extrapolation.
There are fewer DMR frequencies (3) than microwave emission components (4);
consequently, the linear combination method 
still requires correction for dust and synchrotron emission.
The linear combination method identifies signals 
with a specified frequency dependence,
but at the cost of a significant increase in the instrument noise;
the increase in noise can be larger than the amplitude
of the Galactic emission the technique is designed to identify.
The cross-correlation technique removes all emission traced by 
the given template, regardless of frequency dependence,
but does not remove emission uncorrelated with the template.

The cross-correlation technique has been used to identify all three
Galactic emission components in the DMR sky maps (Kogut et al.\ 1996a).
We assume that the DMR maps are a superposition of CMB emission and
Galactic emission traced by a Galactic template,
$$
\Delta T^{\rm DMR} =  \Delta T^{CMB} ~+ ~\alpha \Delta X^{\rm Gal},
$$
where $\Delta T^{\rm DMR}$ is the antenna temperature
in a DMR map,
$\Delta X^{\rm Gal}$ is the intensity of the Galactic template map
(not necessarily in temperature units)
and the coefficient $\alpha$ converts the units of the Galactic map
to antenna temperature at the DMR frequency.
Figure 1 shows the templates used in this Letter.

We estimate the correlation coefficient $\alpha$
by minimizing
\begin{equation}
\chi^2 ~= \sum_{a,b} 
~(D - \alpha G)_a
~({\bf M}^{-1})_{ab}
~(D - \alpha G)_b,
\label{chisq_eq}
\end{equation}
where
$D$ is a linear function of the DMR map temperatures,
$G$ is a similar function for the Galactic template map,
and ${\bf M}$ is the covariance matrix of the function $D$.
We use three linear functions in equation \ref{chisq_eq}:
the 2-point cross-correlation (Kogut et al.\ 1996a),
the temperature in each map pixel (Hinshaw et al.\ 1996),
or Fourier components in an orthogonal basis (G\'{o}rski et al.\ 1996)
and obtain good agreement between the methods.

Uncertainties in the coefficients $\alpha$ are dominated by 
chance alignment of the CMB anisotropy 
with features in the template maps.
Since the relative orientation of the CMB and Galactic template
is unchanged with frequency,
we may reduce the effects of chance alignment
by simultaneously fitting a Galactic template to all three DMR frequencies,
and requiring that the CMB pattern be invariant in thermodynamic units.
Equation \ref{chisq_eq} is easily generalized to 
fit multiple DMR frequencies
and template maps simultaneously.

\section{Galactic Microwave Emission}
Bennett et al.\ (1992) review several models of synchrotron emission
based on radio surveys
at 408 MHz (Haslam et al.\ 1981)
and 1420 MHz (Reich \& Reich 1988).
The simplest model uses the 408 MHz survey
scaled to higher frequencies using
a spatially invariant spectral index
$\beta_{\rm synch} = -2.75$.
A more realistic model uses the
1420 and 408 MHz surveys to trace spatial variation in $\beta_{\rm synch}$
and accounts for the steepening synchrotron spectrum at higher frequencies
using local measurements of the cosmic-ray electron energy spectrum.
Both models have deficiencies and it is possible that neither template
accurately reflects the distribution of synchrotron emission 
at millimeter wavelengths,
regardless of overall normalization.
The large-scale structure in the 408 MHz survey at $|b| > 20\deg$
is dominated by the North Polar Spur (Loop I),
but this region has steeper spectral index
and becomes increasingly less important at higher frequencies.
(Lawson et al.\ 1987).
The cosmic-ray model accounts for 
both the spatial variation in $\beta_{\rm synch}$
and the steepening of the spectrum at higher frequencies,
but the spatial structure of this model at the DMR frequencies
is dominated by regions of flattened index 
at the southern declination limits of the 1420 MHz survey.
The DMR maps show no evidence for such bright regions,
suggestive instead of sidelobe pickup in the 1420 MHz survey.

We obtain an upper limit to synchrotron emission traced by either template
by comparing the synchrotron templates to the DMR 4-year sky maps.
We convolve the templates with the DMR beam profile,
combine the A and B channels at each DMR frequency to form the (A+B)/2 sum,
excise the Galactic plane,
and cross-correlate each synchrotron template 
with the DMR 31.5, 53, and 90 GHz maps.
Table 1 shows the fitted correlation coefficients 
for the two synchrotron templates
derived using Eq.\ \ref{chisq_eq}
in a maximum-likelihood analysis with the ``brute-force'' pixel basis
(Hinshaw et al. 1996).
We evaluate the three DMR frequencies simultaneously,
$T^{\rm DMR} = [T_{31}, T_{53}, T_{90}]$,
and account for possible cross-talk between the synchrotron and far-IR
templates by fitting both templates simultaneously.
The uncertainties in Table 1 include the errors from
instrument noise, chance alignments, and cross-talk with the far-IR template.
We find no statistically significant correlation between 
the DMR sky maps
and either the 408 MHz survey or the cosmic-ray synchrotron model.
Table 2 shows the {\it rms} fluctuations in antenna temperature
corresponding to the fitted coefficients.
We adopt an upper limit $\Delta T_{\rm synch} < 11 ~\muK$ (95\% confidence)
at 31.5 GHz for emission traced by either synchrotron template.

We also fit the DMR maps to the far-infrared dust emission traced
by the DIRBE 140 \um\ map, from which a model of zodiacal dust emission
has been removed (Reach et al.\ 1995).
Tables 1 and 2 show a statistically significant correlation
between the DMR maps and the DIRBE 140 \um\ map;
we obtain nearly identical results using the DIRBE maps at
100 or 240 \um.
The frequency dependence of the inferred signal in the DMR maps
is well described by a superposition of dust and free-free emission:
$\beta = -1.7 \pm 0.8$ between 31.5 and 53 GHz,
and
$\beta = 0.1 \pm 0.9$ between 53 and 90 GHz.
We fit the {\it rms} DMR and DIRBE signals to emission models of the form
$I_\nu = \tau (\frac{\nu}{\nu_0})^{\beta} B_\nu(T)
	~+ A_{\rm ff}(\frac{\nu}{\nu_0})^{-0.15}$,
i.e., a model with a single dust population 
with enhanced submillimeter emissivity plus free-free emission.
The best fit occurs for 
dust temperature $T = 20.0^{+2.5}_{-4.0}$ K
and emissivity $\beta = 1.5^{+1.1}_{-0.3}$,
with opacity $\tau = (1.2^{+0.7}_{-0.4}) \times 10^{-5}$ (68\% confidence).

Table 3 shows the {\it rms} amplitude of the inferred 
dust and free-free signals, including the quadrupole,
in the high-latitude portion of the DMR maps.
Note that the DMR 53 and 90 GHz channels have nearly equal Galactic emission
from spatially correlated dust and free-free emission.
The amplitude of this emission, however, is small
compared to the {\it rms} CMB anisotropy (Banday et al.\ 1996).
The correlation technique, however, is insensitive to
free-free emission whose spatial distribution 
is uncorrelated with the Galactic template maps.
We place upper limits on the amplitude of the uncorrelated component
by analyzing a linear combination of the DMR maps
designed to be sensitive to free-free emission 
(spectral index -2.15 in units of antenna temperature),
cancel emission with a CMB spectrum, and minimize instrument noise:
\begin{equation}
T_{\rm ff} =
  0.37 \times \frac{1}{2}(T^\prime_{\rm 31A} \pm T^\prime_{\rm 31B})
+ 0.02 \times \frac{1}{2}(T^\prime_{\rm 53A} \pm T^\prime_{\rm 53B})
- 0.47 \times \frac{1}{2}(T^\prime_{\rm 90A} \pm T^\prime_{\rm 90B}),
\label{ff_eq}
\end{equation}
where $T^\prime$ is the antenna temperature in each DMR channel
after subtracting synchrotron and dust emission
using the cosmic-ray and DIRBE models, respectively.
We smooth the maps with a 7\deg ~FWHM Gaussian 
to further reduce the effects of noise,
remove a fitted monopole and dipole, and
compare the variance of the (A+B)/2 sum map to the (A-B)/2 difference map.
We obtain an estimate for the fluctuations 
in free-free antenna temperature at 53 GHz from all sources,
$\Delta T_{\rm ff} = 5.2 \pm 4.2 ~\muK$.
This value compares well with the correlated component
at the same effective smoothing,
$\Delta T_{\rm ff} = 6.8 \pm 2.6 ~\muK$.

We test for free-free emission uncorrelated with the far-IR dust
by removing the correlated free-free component from each DMR channel
prior to forming the free-free linear combination (Eq.\ \ref{ff_eq}).
A power spectrum analysis of this uncorrelated free-free map shows
no statistically significant signal at any $\ell < 30$.
The correlated component must form at least 1/3 of 
the total free-free emission (95\% confidence)
and may form the bulk of this emission.

\section{Quadrupole}
The quadrupole in Galactic coordinates is defined by five components $Q_i$,
\begin{eqnarray}
Q(l, b) & = & Q_1 (3 \sin^2b - 1)/2 ~+
	   ~Q_2 \sin 2b ~\cos l ~+
	   ~Q_3 \sin 2b ~\sin l ~+ \nonumber \\
	&  & Q_4 \cos^2b ~\cos 2l ~+
	   ~Q_5 \cos^2b ~\sin 2l,
\label{quad_def}
\end{eqnarray}
with {\it rms} amplitude
$$
Q^2_{rms} = \frac{4}{15}[
	~\frac{3}{4} Q_1^2 +
	~Q_2^2 + ~Q_3^2 + ~Q_4^2 + ~Q_5^2 ~].
$$
Note that $Q_{rms}$ refers to the {\it rms} amplitude of the observed 
quadrupole in the sky, while $Q_{rms-PS}$ is a model parameter 
giving the CMB power spectrum normalization; the two are equivalent
only in an ensemble average over many Hubble volumes.

The quadrupole represents one order of a spherical harmonic expansion
of the temperature distribution in the sky maps.  
After the low-latitude portion of the sky map is excised,
the spherical harmonic functions $Y_{\ell m}(l,b)$
are no longer orthogonal on the remaining sky pixels,
allowing power from higher orders to be aliased into the quadrupole
(and vice versa).
G\'{o}rski et al.\ (1996) address this problem by constructing a new set of
basis functions orthogonal on the cut sky.
In what follows, we retain the more familiar quadrupole basis
and minimize aliasing by including a theoretical model for the
higher orders ($\ell \ge 3$) of the power spectrum.
We derive the quadrupole parameters $Q_i$ by minimizing
\begin{equation}
\chi^2 = \sum_{ab} 
	~[ ~T^{\rm DMR} - (\sum_i Q_i B_i) ]_a
	~({\bf M}^{-1})_{ab}
	~[ ~T^{\rm DMR} - (\sum_i Q_i B_i) ]_b,
\label{quad_eq}
\end{equation}
in a pixel-based maximum-likelihood analysis, where
$B_i$ are the quadrupole basis functions from Eq.\ \ref{quad_def}
and the $954 \times 954$ covariance matrix ${\bf M}$ 
is defined for a scale-invariant CMB model
({\it cf.} Eq.\ 1 of Hinshaw et al.\ 1996).
That is, we describe the high-latitude sky as a fixed quadrupole pattern
plus a statistical distribution of higher-order power
given by a Harrison-Zel'dovich power spectrum.
Table 4 shows the quadrupole parameters $Q_i$ fitted to the 
high-latitude portion of the (A+B)/2 sky maps, 
without Galactic correction.
The quadrupole parameters for the Galactic template maps are also shown.
The $Q_i$ in Table 4 are for a single Galactic cut.
We have repeated the analysis for a variety of cut angles.
Not unexpectedly, the fitted values show a strong dependence on latitude $|b|$:
the Galaxy is a strong quadrupolar source
(see, e.g., Table 2 of Bennett et al.\ 1992).

We derive the CMB quadrupole parameters as follows.
We correct each DMR map for the second-order Doppler quadrupole
$[Q_1, Q_2, Q_3, Q_4, Q_5] = [0.9, -0.2, -2.0, -0.9, 0.2] ~\muK$
thermodynamic temperature
caused by the motion of the solar system with respect to the CMB rest frame.
We remove Galactic emission using the synchrotron and DIRBE templates
by adding terms to Eq.\ \ref{quad_eq} of the form
$\alpha_{\rm synch} X^{\rm synch} \nu^{\beta_{\rm synch}} ~+
~\alpha_{\rm dust} X^{\rm DIRBE}  \nu^{\beta_{\rm dust}}~+
~\alpha_{\rm ff} X^{\rm DIRBE} \nu^{\beta_{\rm ff}} $
and fitting the three DMR frequencies simultaneously.
Rather than fitting the synchrotron and DIRBE templates 
to each map independently,
we use the results of {\S}3
and fit the templates to all three frequencies simultaneously,
specifying the spectral behavior
via the spectral indices.
The free parameters in the fit are thus
the synchrotron, dust, and free-free amplitude coefficients
plus 5 CMB quadrupole parameters.
Using the nominal values
$\beta_{\rm synch} = -3, ~\beta_{\rm dust} = +2,$ 
and $\beta_{\rm ff} = -2.15$,
we derive correlation coefficients
$\alpha_{\rm synch} = 0.27 \pm 0.20 ~\muK ~{\rm K}^{-1}$,
$\alpha_{\rm dust} =  0.70 \pm 0.27 ~\muK ~{\rm MJy}^{-1} ~{\rm sr}$, and
$\alpha_{\rm ff} =    2.07 \pm 0.49 ~\muK ~{\rm MJy}^{-1} ~{\rm sr}$
antenna temperature at 53 GHz.
Table 5 shows the quadrupole parameters for the corrected CMB map.
The uncertainties are dominated by 
aliasing of higher-order power
and to a lesser extent by instrument noise
and chance alignments in the template map correlation.

The statistical uncertainties in Table 5 
do not include the systematic effects associated with
the choice of template map and spectral index.
We evaluate the uncertainties associated with the Galactic spectral indices
by repeating the analysis as
$\beta_{\rm synch}$ is varied over the range [-2.8, -3.3]
and $\beta_{\rm dust}$ over the range [1.1, 2.0].
The differences are small, typically 10\% of the statistical uncertainty.

We estimate the uncertainties 
associated with the Galactic model techniques
by repeating the CMB quadrupole analysis 
for several different Galactic models.
The simplest change is to substitute the cosmic-ray synchrotron template
for the 408 MHz survey.  
Synchrotron emission is faint at millimeter wavelengths,
and both templates have similar quadrupoles:
the estimated CMB quadrupole does not depend sensitively 
on the choice of synchrotron template or spectral index.
Similarly, we may estimate the dust emission by fitting the
DIRBE dust emission or scaling the FIRAS dust model
(Wright et al.\ 1991).
Differences between these techniques are small
and are dominated by differences in the FIRAS and DMR beam shape.

We estimate the systematic uncertainty in the free-free correction
by comparing the quadrupole results using the DMR/DIRBE cross-correlation
to the results using the linear combination technique.
For the latter analysis,
we correct the DMR channel maps for dust and synchrotron emission
and form a weighted CMB map using weights
-0.195, -0.115, +0.394, +0.286, +0.313, and +0.507
for antenna temperature in channels 
31A, 31B, 53A, 53B, 90A, and 90B, respectively.
This linear combination cancels free-free emission (spectral index -2.15)
with maximal instrument sensitivity,
although the resulting noise is still 35\% larger 
than the noise using the cross-correlation technique.
We then fit for the CMB quadrupole using Eq.\ \ref{quad_eq},
replacing $T^{\rm DMR}$ with the linear combination CMB map.
The results are in agreement with the less noisy cross-correlation technique
(Table 5).
Systematic effects associated with changing the Galactic model
are typically 1--5 \muK, smaller than the statistical uncertainties.

The largest systematic effect is the choice of Galactic cut angle.
Table 5 shows the quadrupole parameters fitted for
the regions $|b| > 15\deg$, $|b| > 20\deg$ with custom cutouts,
and $|b| > 30\deg$.
The changes in the fitted parameters as the Galactic cut is varied
are comparable to the statistical uncertainties,
and limit our ability to estimate the CMB quadrupole.
Since statistically significant CMB features exist in the region
$15\deg \lt b \lt 30\deg$,
the change in fitted quadrupole parameters as the Galactic cut is varied
reflects aliasing of higher-order power as well as potential
shortcomings in the Galactic models.
We estimate the uncertainty from this effect as 
half of the spread in each quadrupole component.
The bottom row of Table 5 shows the combined systematic uncertainty
in each component,
given by the quadrature sum of the separate uncertainties
in spectral index, model techniques, Galactic cut,
and the instrument systematic artifacts (Kogut et al.\ 1996b).

The {\it rms} quadrupole amplitude, $Q_{rms}$, 
provides a convenient description of the CMB quadrupole.
Since it is quadratic in the map temperatures,
both instrument noise and aliasing will bias $Q_{rms}$ toward higher values
(the individual parameters $Q_i$ are unbiased estimators, but provide
a less concise characterization).
We correct for the bias by the quadrature subtraction of the 
{\it rms} statistical uncertainty:
$ Q^2_{rms} = \frac{4}{15}
(~\sum_i \eta_i Q_i^2 - ~\sum \eta_i \delta Q_i^2 ~)$,
where $Q_i$ are the central values of the fitted quadrupole components,
$\delta Q_i$ are the statistical uncertainties,
and $\eta_i$ are the weights [$\frac{3}{4}$, 1, 1, 1, 1]
(Gould 1993).
Column 6 of Table 5 (labelled $Q_{rms}$) shows this de-biased estimate
for the CMB {\it rms} quadrupole amplitude.

Figure 2 shows the CMB maps from the 
cross-correlation and linear combination techniques.
Since the cross-correlation technique has the smallest random errors,
we adopt the CMB quadrupole parameters from this model
using the custom Galactic cut
as a compromise between 
sky coverage and Galactic foreground emission.
The resulting estimate of the CMB quadrupole amplitude is
$Q_{rms} = \MeanQ \pm \dQStat \pm \dQGal ~\muK$,
where errors represent the statistical and systematic uncertainties,
respectively.
Allowing for the systematic uncertainty in the Galactic correction,
the CMB quadrupole amplitude lies in the range 
[4, 28] \muK\ at 95\% confidence.

\section{Conclusions}
We use the DMR 4-year sky maps and Galaxy-dominated sky surveys
to derive models of high-latitude Galactic and cosmic emission.
Cross-correlation of the DMR maps with either
the 408 MHz synchrotron survey or a cosmic-ray synchrotron model
with spatially varying spectral index 
yield upper limits to fluctuations in synchrotron emission 
traced by either template,
$\Delta T_{\rm synch} < 11 ~\muK$ at 31.5 GHz.
If either template correctly reproduces the angular distribution of synchrotron 
emission, the amplitude normalization requires a mean spectral index
between 408 MHz and 31.5 GHz of $\beta_{\rm synch} < -3.0$.

Cross-correlation of the DMR maps 
with the dust-dominated DIRBE 140 \um\ survey
shows a statistically significant signal
whose dependence on DMR frequency is consistent with a superposition
of dust and free-free emission.
The {\it rms} amplitude of the dust signal is
$\Delta T_{\rm dust} = 2.7 \pm 1.3 ~\muK$
antenna temperature at 53 GHz,
including the contribution from the quadrupolar component.
The dust emission may be used to place a lower limit to
enhanced emissivity from 140 $\mu$m to 6 mm wavelenth:
$\beta_{\rm dust} > 1.1$ at 95\% confidence.

We detect a component of the free-free emission from the warm ionized 
interstellar medium that is correlated with the far-infrared dust
on angular scales of 7\deg ~or larger.
The amplitude of the correlated free-free emission,
$\Delta T_{\rm ff} = 7.1 \pm 1.7 ~\muK$,
compares well with the total free-free emission from all sources,
derived from a linear combination of the DMR maps.
The correlated component must form at least 1/3 of 
the total free-free emission (95\% confidence)
and may form the bulk of this emission.

Galactic emission is comparable to the CMB on quadrupolar scales
and is counter-aligned to the CMB in several of the 5 components.
The fitted quadrupole in the uncorrected DMR maps is not representative
of the CMB quadrupole in either amplitude or phase.
Analysis of the DMR maps that includes the quadrupole anisotropy
should correct for Galactic emission,
using either the cross-correlation technique
or the noisier linear combination technique.
We correct the DMR channel maps for Galactic emission
using both techniques
and estimate the quadrupole components
using a maximum likelihood analysis.
The random statistical uncertainties are dominated 
by the aliasing of power from higher multipole orders.
After correction for the positive bias from noise and aliasing,
the CMB quadrupole amplitude observed at high latitude is
$Q_{rms} = \MeanQ \pm \dQStat \pm \dQGal ~\muK$,
where the uncertainties represent the random statistical errors
and systematic modelling errors,
respectively.

\vspace{0.5 in}
We are grateful to Charley Lineweaver for helpful discussion.
We acknowledge the dedicated efforts
of the many people responsible for the {\it COBE} DMR data:
the NASA office of Space Sciences,
the {\it COBE} flight team,
and all of those who helped process and analyze the data.

%
\vfill
\clearpage

\normalsize
\halfspace
\begin{table}
\caption{DMR-Galactic Template Cross-Correlation Coefficients$^a$}
\begin{center}
\begin{tabular}{l c c c}
\hline
DMR Frequency	& \multicolumn{3}{c}{Galactic Template} \\
(GHz)		& 408 MHz$^b$ & Cosmic-Ray$^c$ & DIRBE 140 \um$^d$ \\
\hline
31.5		&  1.17 $\pm$ 1.13 & 1.88 $\pm$ 1.24 & 6.37 $\pm$ 1.52 \\
53		&  0.69 $\pm$ 0.77 & 0.88 $\pm$ 0.81 & 2.69 $\pm$ 1.06 \\
90		& -0.14 $\pm$ 0.74 & 0.43 $\pm$ 2.55 & 2.79 $\pm$ 1.01 \\
\hline
\end{tabular}
\end{center}
$^a$ The antenna temperature at each DMR frequency
of Galactic emission traced by template map X
is $T_A = \alpha X$ \muK.
Results are quoted for $|b| > 20\deg$ 
with custom cutouts at Orion and Ophiuchus (see text). \\
$^b ~\alpha$ has units \muK ~K$^{-1}$ since the template map has units K.\\
$^c ~\alpha$ is dimensionless since the template map has units \muK.\\
$^d ~\alpha$ has units \muK ~(MJy/sr)$^{-1}$ since the template map
has units MJy sr$^{-1}$.
\end{table}

\normalsize
\halfspace
\begin{table}
\caption{RMS Galactic Signal in DMR Sky Maps (\muK)$^a$}
\begin{center}
\begin{tabular}{l c c c}
\hline
DMR Frequency	& \multicolumn{3}{c}{Galactic Template} \\
(GHz)		& 408 MHz & Cosmic-Ray & DIRBE 140 \um \\
\hline
31.5	&  5.7 $\pm$ 5.5  &  8.4 $\pm$ 5.5  &  22.7 $\pm$ 5.4 \\
53	&  3.4 $\pm$ 3.7  &  3.9 $\pm$ 3.6  &   9.6 $\pm$ 3.8 \\
90	& -0.7 $\pm$ 3.6  &  1.9 $\pm$ 3.5  &  10.0 $\pm$ 3.6 \\
\hline
\end{tabular}
\end{center}
$^a$ Units are antenna temperature.  The quadrupole has not been subtracted.
Results are quoted for $|b| > 20\deg$ 
with custom cutouts at Orion and Ophiuchus. \\
\end{table}

\normalsize
\halfspace
\begin{table}
\caption{Correlated Dust and Free-Free Antenna Temperature$^a$}
\begin{center}
\begin{tabular}{l c c c}
\hline
DMR Frequency	&  {\it rms} Free-Free	& {\it rms} Dust \\
(GHz)		&  (\muK)		&  (\muK) \\
\hline
31.5 		&  21.7 $\pm$ 5.2	& 0.9 $\pm$ 0.4 \\
53 		&   7.1 $\pm$ 1.7	& 2.7 $\pm$ 1.3 \\
90 		&   2.3 $\pm$ 0.5	& 7.6 $\pm$ 3.6 \\
\hline
\end{tabular}
\end{center}
$^a$ Results are quoted for $|b| > 20\deg$ 
with custom cutouts at Orion and Ophiuchus. \\
\end{table}

\normalsize
\halfspace
\begin{table}
\caption{Quadrupole Components in Uncorrected Sky Maps$^a$}
\begin{center}
\begin{tabular}{l c c c c c c}
\hline
Map & $Q_1$ & $Q_2$ & $Q_3$ & $Q_4$ & $Q_5$ & $Q_{rms}^b$ \\
\hline
DMR 31.5 GHz$^c$ & 
-29.1 $\pm$ 12.6 &  27.0 $\pm$  6.5 &   2.1 $\pm$  6.0 & 
  4.5 $\pm$ 12.8 &   5.3 $\pm$ 11.9 &  15.6 $\pm$  5.4 \\
DMR 53 GHz$^c$ & 
-13.6 $\pm$  7.7 &   9.5 $\pm$  2.7 &   3.8 $\pm$  2.2 & 
  0.0 $\pm$  8.2 &   3.5 $\pm$  7.6 &   4.4 $\pm$  3.3 \\
DMR 90 GHz$^c$ & 
  0.2 $\pm$  8.5 &   5.0 $\pm$  3.5 &   6.7 $\pm$  2.8 & 
 -0.3 $\pm$  9.1 &   3.2 $\pm$  8.3 &   0.0 $\pm$  3.0 \\
408 MHz$^d$        & -6.6 & 3.4 & -0.1 & 5.5 & -1.8 & 4.5 \\
Cosmic-ray$^e$     & -5.2 & 2.7 & -1.3 & 5.9 & -1.3 & 4.2 \\
DIRBE 140 \um$^f$  & -6.6 & 1.7 & -1.3 & 1.9 & -1.4 & 3.4 \\
\hline
\end{tabular}
\end{center}
$^a$ All results for high-latitude sky, $|b| > 20\deg$ with
custom cutouts at Orion and Ophiuchus.  The Doppler quadrupole
has been removed from the DMR maps. \\
$^b$ $Q_{rms}$ has been corrected for statistical bias (see text). \\
$^c$ Map units are \muK ~thermodynamic temperature. \\
$^d$ Map units are K antenna temperature. \\
$^e$ Map units are \muK ~antenna temperature. \\
$^f$ Map units are MJy sr$^{-1}$.
\end{table}

\normalsize
\halfspace
\begin{table}
\caption{CMB Quadrupole Parameters (\muK)$^a$}
\begin{center}
\begin{tabular}{c c c c c c}
\hline
$Q_1$ & $Q_2$ & $Q_3$ & $Q_4$ & $Q_5$ & $Q_{rms}^b$ \\
\hline
\multicolumn{6}{l}{Cross-Correlation} \\
 27.9 $\pm$  4.7 &   0.6 $\pm$  1.7 &  10.8 $\pm$  1.6 & 
-14.9 $\pm$  4.9 &  19.0 $\pm$  4.6 &  18.0 $\pm$  2.3 \\
 19.0 $\pm$  7.4 &   2.1 $\pm$  2.5 &   8.9 $\pm$  2.0 & 
-10.4 $\pm$  8.0 &  11.7 $\pm$  7.3 &  10.7 $\pm$  3.6 \\
 31.2 $\pm$ 12.5 &   1.0 $\pm$  3.7 &  12.8 $\pm$  3.7 & 
 -9.3 $\pm$ 13.4 &  -0.6 $\pm$ 12.1 &  11.7 $\pm$  5.8 \\
 & & & & & \\
\multicolumn{6}{l}{Linear Combination} \\
 24.3 $\pm$  6.2 &  -2.1 $\pm$  3.1 &   8.3 $\pm$  2.9 & 
-18.9 $\pm$  6.2 &   6.8 $\pm$  6.0 &  14.6 $\pm$  3.1 \\
 20.0 $\pm$  9.1 &  -2.6 $\pm$  3.9 &   9.4 $\pm$  3.3 & 
 -6.1 $\pm$  9.6 &   3.5 $\pm$  8.8 &   7.1 $\pm$  4.2 \\
 35.1 $\pm$ 14.3 &  -3.2 $\pm$  4.9 &  12.7 $\pm$  4.8 & 
 -7.1 $\pm$ 15.4 &   6.6 $\pm$ 13.8 &  12.2 $\pm$  6.8 \\
 & & & & & \\
\multicolumn{6}{l}{Systematic and Model Uncertainties} \\
8.2 & 2.7 & 2.5 & 4.3 & 10.4 & 7.1 \\
\hline
\end{tabular}
\end{center}
$^a$ The three rows for each technique refer to Galactic cut
$|b| > 15\deg$, $|b| > 20\deg$ with custom cutouts, and $|b| > 30\deg$.
The Doppler quadrupole has been removed from the maps.
Results are in thermodynamic temperature. \\
$^b$ $Q_{rms}$ has been corrected for statistical bias (see text).
\end{table}
 
\clearpage
\begin{center}
\large
{\bf References}
\end{center}

\normalsize
\pprintspace

\refitem
Banday, A.\ \& Wolfendale, A.W.\ 1991, MNRAS, 248, 705

\refitem
---, et al.\ 1996, ApJ Letters, in preparation

\refitem
Bennett, C.L., et al.\ 1992, ApJ, 396, L7

\refitem
---, et al.\ 1994, ApJ, 434, 587

\refitem
---, et al.\ 1996, ApJ Letters, submitted

\refitem
Bensadoun, M., Bersanelli, M., De Amici, G., Kogut, A., Levin, S.M., Limon, M., 
Smoot, G.F., \& Witebsky, C.\ 1993, ApJ, 409, 1

\refitem
D\'{e}sert, F.-X., Boulanger, F., \& Puget, J.-L.\ 1990, A\&A, 327, 215

\refitem
Fixsen, D.J., Cheng, E.S., \& Wilkinson, D.T.\ 1983, PRL, 50, 620

\refitem
Gautier, T.N., Boulanger, F., P\'{e}rault, M., \& Puget, J.L.\ 1992,
AJ, 103, 1313

\refitem
G\'{o}rski, K.M., Banday, A.J., Bennett, C.L., Hinshaw, G., Kogut, A.,
Smoot, G.F., \& Wright, E.L.\ 1996, ApJ Letters, submitted
1996, ApJ Letters, submitted

\refitem
Gould, A.\ 1993, ApJ, 403, L51

\refitem
Guti\'{e}rrez de la Cruz, C.M., Davies, R.D., Rebolo, R., Watson, R.A., 
Hancock, S., \& Lasenby, A.N.\ 1995, ApJ, 442, 10

\refitem
Haslam, C.G.T, Klein, U., Salter, C.J., Stoffel, H, Wilson, W.E., Cleary, M.N., 
Cooke, D.J., \& Thomasson, P.\ 1981, A\&A, 100, 209

\refitem
Hinshaw, G., Banday, A.J., Bennett, C.L., G\'{o}rski, K.M.,
Kogut, A., Smoot, G.F., \& Wright, E.L.\ 1996, ApJ Letters, submitted

\refitem
Kogut, A., Banday, A.J., Bennett, C.L., G\'{o}rski, K.M., 
Hinshaw, G., \& Reach, W.T.\ 1996a, ApJ, 460, in press	

\refitem
---, et al.\ 1996b, ApJ, submitted	

\refitem
Lawson, K.D., Mayer, C.J., Osborne, J.L, \& Parkinson, M.L.\ 1987, 
MNRAS, 225, 307

\refitem
Lubin, P., Villela, T., Epstein, G., \& Smoot, G.\ 1985, ApJ, 298, L1

\refitem
Reach, W.T., Franz, B.A., Kelsall, T., \& Weiland, J.L.\ 1995, 
{\it Unveiling the Cosmic Infrared Background}, ed.\ E.\ Dwek, (New York:AIP)

\refitem
Reich, P., \& Reich, W.\ 1988, A\&AS, 74, 7

\refitem
Reynolds, R.J.\ 1992, ApJ, 392, L35

\refitem
Wright, E.L., et al.\ 1991, ApJ, 381, 200

\clearpage
\begin{figure}[t]
\epsfxsize=6.0truein
\epsfbox{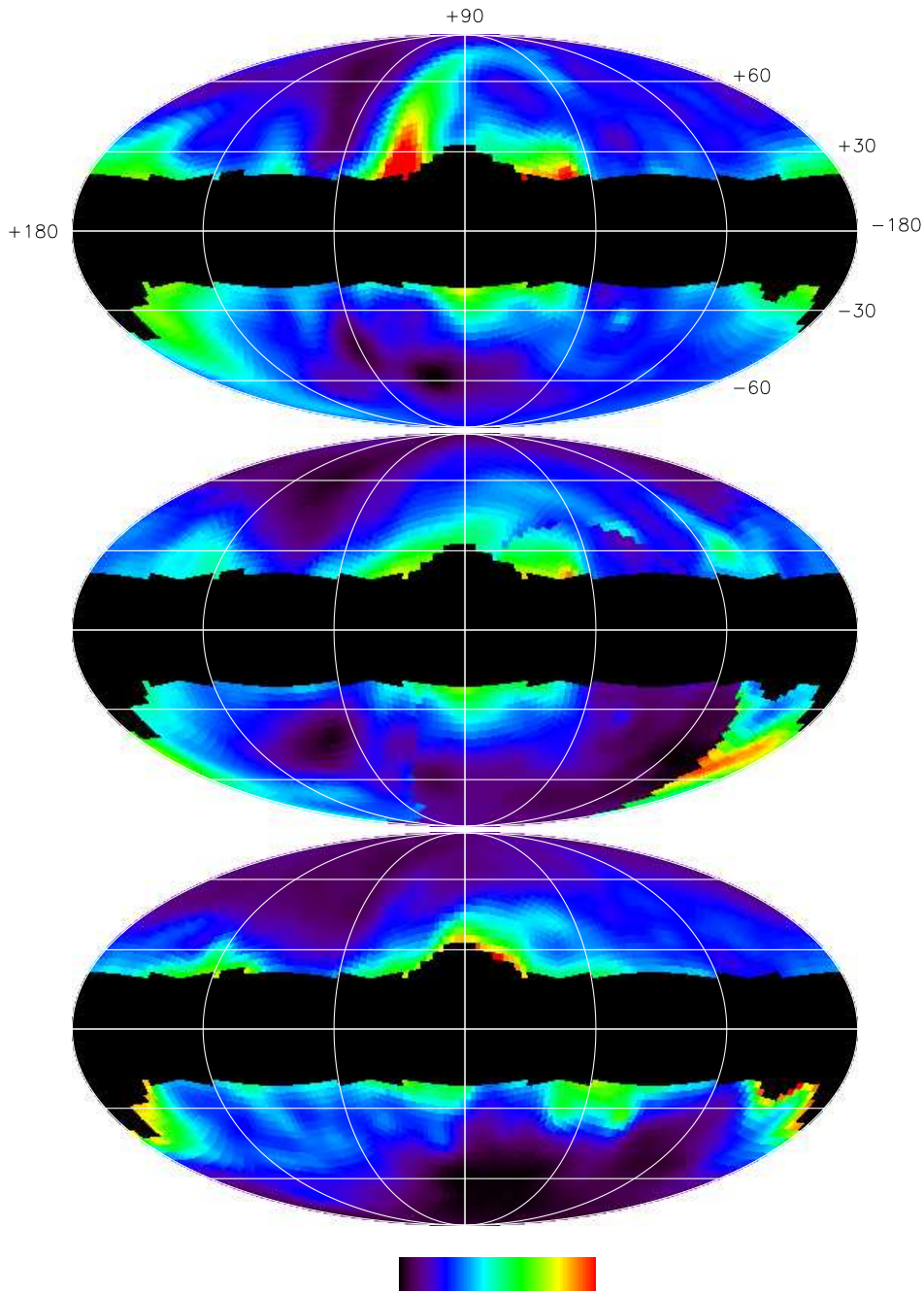}
\caption{
Galactic templates used in cross-correlation technique,
Mollweide projection in Galactic coordinates.
(Top) Synchrotron-dominated 408 MHz survey.
(Middle) Cosmic-ray synchrotron model.
(Bottom) Dust-dominated DIRBE 140 \um\ survey.
Each template has been convolved with the DMR beam pattern and 
is masked to show only data at high Galactic latitudes;
a fitted monopole and dipole have been removed from the remaining pixels.
The color transfer for each template runs 6 standard deviations
from the coldest pixel.
}
\end{figure}

\clearpage
\begin{figure}[t]
\epsfxsize=6.0truein
\epsfbox{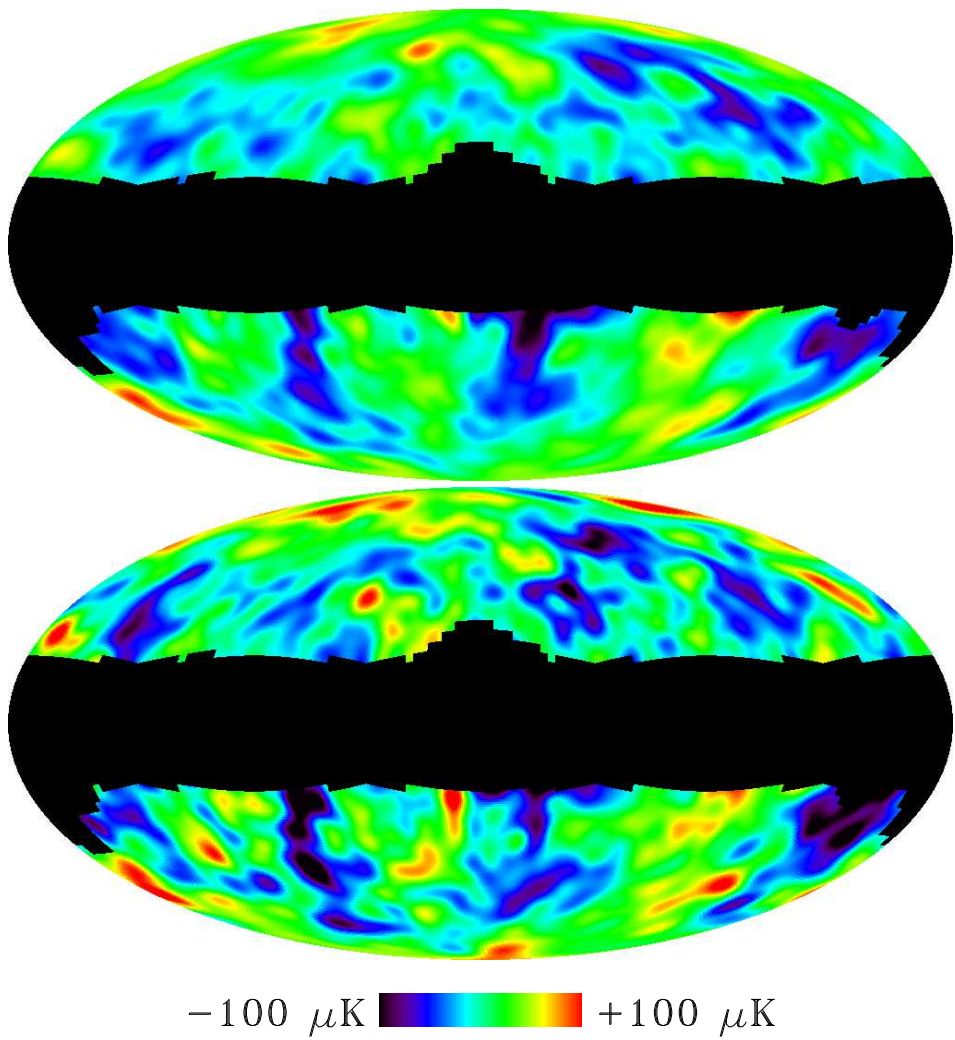}
\caption{
Maps of the cosmic microwave background anisotropy
after removing Galactic emission,
Mollweide projection in Galactic coordinates.
(Top) Cross-correlation technique.
(Bottom) Linear combination technique.
The Galactic models do not attempt to fit the Galactic plane;
only the high-latitude portion is shown.
The cross-correlation technique represents a compromise
between removing faint Galactic emission
and minimizing instrument noise.
}
\end{figure}

\end{document}